# Giant superconducting diode effect in ion-beam patterned Sn-based superconductor nanowire / topological Dirac semimetal planar heterostructures

Keita Ishihara[1,†], Le Duc Anh[1,2,3,*,†], Tomoki Hotta[1], Kohdai Inagaki[1], Masaki Kobayashi[1,3], and Masaaki Tanaka[1,3,4,*]

[1] *Department of Electrical Engineering and Information Systems, The University of Tokyo, Bunkyo-ku, Tokyo 113-8656, Japan.*
[2] *PRESTO, Japan Science and Technology Agency, Kawaguchi, Saitama, 332-0012, Japan*
[3] *Centre for Spintronics Research Network, The University of Tokyo, Bunkyo-ku, Tokyo 113-8656, Japan*
[4] *Institute for Nano Quantum Information Electronics, The University of Tokyo, Komaba, Meguro-ku, Tokyo 153-8505, Japan*

[*] Email: anh@cryst.t.u-tokyo.ac.jp
masaaki@ee.t.u-tokyo.ac.jp
†These authors equally contributed to this work.

**Superconductor/topological material heterostructures are intensively studied as a platform for topological superconductivity and Majorana physics[1-15]. However, the high cost of nanofabrication and the difficulty of preparing high-quality interfaces between the two dissimilar materials are common obstacles that hinder the observation of intrinsic physics and the realisation of scalable topological devices and circuits. Here, we demonstrate an innovative method to directly draw nanoscale superconducting beta-tin (β-Sn) patterns of any shape in the plane of a topological Dirac semimetal (TDS) alpha-tin (α-Sn) thin film[16] by irradiating a focused ion beam (FIB). We utilise the property that α-Sn undergoes a phase transition to superconducting β-Sn upon heating by FIB. In β-Sn nanowires embedded in a TDS α-Sn thin film, we observe giant non-reciprocal superconducting transport, where the critical current changes by 69% upon reversing the current direction. The superconducting diode rectification ratio $\eta$ reaches a maximum when the magnetic field is applied parallel to the current,**

**distinguishing itself from all the previous reports. Moreover, it oscillates between alternate signs with increasing magnetic field strength. The angular dependence of $\eta$ on the magnetic field and current directions is similar to that of the chiral anomaly effect in TDS α-Sn, suggesting that the non-reciprocal superconducting transport may occur at the β-Sn/α-Sn interfaces. The ion-beam patterned Sn-based superconductor/TDS planar structures thus show promise as a universal platform for investigating novel quantum physics and devices based on topological superconducting circuits of any shape.**

Topological superconductors garner significant attention as a crucial pathway for realizing Majorana quasiparticles, which are highly sought-after yet enigmatic candidates for quantum bits in fault-tolerant quantum computing[1–4]. To date, extensive research on topological superconductivity has focused on heterostructures of topologically nontrivial materials, such as topological insulators[5,6], TDS[7,8], and topological Weyl semi-metals[9,10], in conjunction with conventional s-wave superconductors, exploiting the superconducting proximity effect[5–15]. To achieve a material platform for large-scale quantum computing, the critical challenges in such heterostructures are establishing high-quality interfaces between topological materials and superconductors while utilizing cost-effective and scalable processes for material growth and nanofabrication. Given that many candidate topological materials are relatively new and immature, meeting these requirements presents a significant challenge for materials science and device engineering.

This work introduces a novel topological material platform based on Sn, one of human history's earliest-known and most commonly used metal elements. Body-centered tetragonal β-Sn is a well-known metal that exhibits conventional BCS-type superconductivity below 4 K. On the other hand, α-Sn, with a diamond-type crystal structure, behaves as a Luttinger semimetal with no band gap in bulk. However, under tensile or compressive strain, it can transform into a three-dimensional topological insulator (3D-TI) or a TDS, respectively. Notably, TDSs[16–18] feature 3D linear-dispersion Dirac cones with strong spin-orbit-momentum locking, leading to exotic phenomena such as Fermi arcs on the surface and chiral anomaly. Both the 3D Luttinger semimetal and TDS have been theoretically predicted to become topological superconductors capable of hosting Majorana quasiparticles in their superconducting states[19,20].

Among the rare examples of experimentally confirmed TDSs[16-18,21-27], α-Sn [16,23-26] stands out as the only elemental material that can be grown with exceptional quality on semiconductor substrates like InSb or CdTe. Furthermore, α-Sn undergoes a phase transition to β-Sn upon heating, providing a new avenue to introduce superconductivity into the already-rich topological phase diagram of α-Sn. This study presents an innovative method to incorporate superconductivity into the topological platform of α-Sn, forming arbitrary planar shape patterns on the nanometer scale. Using a Ga-FIB, we directly draw nm-scale β-Sn patterns onto TDS α-Sn thin films, inducing a phase transition from α-Sn to superconducting β-Sn upon FIB irradiation. While previous reports have utilised ion beams to partially destroy superconductors and create insulating layers for fabricating Josephson junctions[28-31], our approach results in superconducting β-Sn regions within the TDS α-Sn films with high-quality interfaces. As will be shown in this work, β-Sn superconducting nanowires embedded in α-Sn exhibit remarkable phenomena. When a moderate in-plane magnetic field is applied, we observe: i) unconventional superconductivity with a two-fold symmetric angular dependence of the critical current ($I_C$), and ii) giant non-reciprocal superconducting transport, where the critical current changes by 69% upon reversing the current direction when the magnetic field is parallel to the current. These unprecedented observations suggest that the combination of TDS and superconductivity in Sn plays a pivotal role in accessing new topological superconducting functionalities. By leveraging these FIB-patterned Sn-based planar structures, we can establish a versatile superconductor/topological material platform for high-quality superconducting circuits of arbitrary shapes; all achieved through remarkably simple processes.

We epitaxially grew 70 nm-thick α-Sn thin films on InSb (001) substrates using molecular-beam epitaxy (see Methods). The growth conditions were the same as our previous study [16], where α-Sn exhibited TDS behaviour due to compressive strain in the in-plane direction. Subsequently, the α-Sn thin film was partially irradiated with a FIB, employing a Ga ion beam accelerated with a voltage of 30 kV and a pitch interval of 5 nm in a serpentine scanning mode (Fig. 1a). The irradiated areas of TDS α-Sn were immediately transformed into β-Sn, easily distinguishable by its metallic silver colour in contrast to the dark grey colour of the surrounding α-Sn regions. We conducted density of states (DOS) measurements at the Fermi level ($E_F$) for both the as-grown α-Sn and the FIB-irradiated Sn using X-ray photoemission spectroscopy (XPS). The results display a clear difference, as illustrated in Fig. 1b: The as-grown α-Sn exhibits vanishingly small DOS at $E_F$, typical of a semi-metal, while the irradiated Sn area shows a much larger DOS with a Fermi-Dirac edge, similar to that of a reference Au thin film, indicating the expected metallic band structure of β-Sn.

For further characterisation of the α-Sn/β-Sn planar structures, we employed microscopic techniques such as scanning electron microscopy (SEM) and scanning transmission electron microscopy (STEM). As depicted in Fig. 1a, top-view SEM images illustrate two typical structures: a β-Sn nanowire embedded in α-Sn (bottom left) and a planar β-Sn/α-Sn/β-Sn junction (bottom right). The β-Sn nanowire, fabricated by irradiating Ga ions with a designed line width $W_0$ of 10 nm, exhibits an actual width $W$ of 180 nm. It is important to note that the β-Sn wire width $W$ is approximately 170 nm larger than the irradiated width $W_0$, owing to the lateral diffusion of heat generated by the Ga ion collision, which additionally induced a phase transition in the α-Sn on both sides. This scenario is confirmed through the cross-sectional STEM image of a nanowire shown in Fig. 1c (see also Extended Data Fig. 1): The β-Sn region consists of a center part with grain boundaries, where the Ga ions directly collided and two clean polycrystalline parts extending 150 – 200 nm on each side, likely formed due to thermal diffusion. The α-Sn/β-Sn interface is found to be atomically abrupt and of high quality. Taking into account the heat-induced expansion of the β-Sn area, we successfully fabricated a planar β-Sn/α-Sn/β-Sn junction with the actual width of the α-Sn channel measuring 70 nm (Fig. 1b), using a designed value of 200 nm. As a result, through simulation and optimizing the patterning design, our FIB method enables the fabrication of any nanoscale planar structures of α-Sn and β-Sn.

Next, we investigated the superconducting properties of the α-Sn/β-Sn planar heterostructures. Three β-Sn nanowires with actual widths $W$ = 500, 1000, and 2000 nm were fabricated along the $[1\bar{1}0]$ axis on a single α-Sn film, all of which exhibited superconductivity below 4 K, as illustrated in Fig. 1d. The resistance displays an anomalous rise just before dropping to zero at the superconducting transition around 3.7 K. This behaviour can be attributed to a decrease in the DOS near $E_F$ when Cooper pairs are formed in the α-Sn due to the superconducting proximity effect at the α-Sn/β-Sn interface[32]. Interestingly, we observe a giant superconducting diode effect (SDE) in all three β-Sn nanowires when a magnetic field is applied parallel and antiparallel to the current at 2 K. Figure 2a shows the resistance – current characteristics in the β-Sn nanowire with $W$ = 500 nm (which we name NW90). When a magnetic field **H** = ±0.11 T is applied parallel to the nanowire, the critical current $I_C$ increases when the current **I** and **H** are parallel ($I_{C+}$) and decreases significantly when **I** and **H** are antiparallel ($I_{C-}$). As shown in Figure 2a, $I_C$ = 61 μA at $H$ = 0 (black curve), $I_{C+}$ = –71 μA and $I_{C-}$ = 29 μA at $H$ = 0.11 T (green curve), thus the change of $I_C$ (= $|I_{C+}| - |I_{C-}|$) is as large as 69% of the $I_C$ at $H$ = 0, which is among the largest SDEs reported thus far. The superconducting diode operation is demonstrated by switching the current at ± 68 μA under a fixed **H** = 0.11 T, as shown in Fig. 2b. The nanowire's resistance alternates between zero and a finite value of 1.6 Ω, indicating a successful demonstration of a superconducting diode operation. The diode efficiency $\eta = ||I_{C+}| - |I_{C-}||/(|I_{C+}| + |I_{C-}|)$ becomes finite and increases monotonically as $H$ is increased, reaching a maximum value of 35% around 1T, which is also one of the largest values of η reported so far, then decreases monotonically to zero at around 0.2 T (Fig. 2c).

SDE has been reported in many recent material systems, and its mechanisms are subjects of active debates[33-46]. In all previous systems, SDE occurs when an applied magnetic field (**H**) is orthogonal to the current (**I**), regardless of whether the proposed mechanisms are intrinsic[33-44] or extrinsic[45,46]. However, the non-reciprocal superconductivity in our β-Sn wires embedded in α-Sn represents the first instance of giant SDE realised when **H** and **I** are *parallel*, a completely different result. To elucidate the new mechanism in the Sn-based nanowires, we investigate the dependence of the superconducting critical current $I_C$ ($I_{C+}$, $I_{C-}$) on the directions of **H** and **I**. As depicted in Figure 3, we patterned three β-Sn nanowires (NW0, NW45, NW90) with the same width ($W = 500$ nm) in a 70 nm-thick α-Sn thin film along three different crystallographic axes of the InSb (001) substrate: [110], [100], and [1$\bar{1}$0], respectively. We measured $I_C$ while rotating the nanowires in a fixed $H$ (= 0.056, 0.185, 0.110 T for NW0, NW45, and NW90, respectively) so that **H** remained in the film plane, all at 2 K. In Figure 3, we show the even component ($I_{EVEN} = (I_{C+} + I_{C-})/2$) and the odd component ($I_{ODD} = |I_{C+} - I_{C-}|/2$) of $I_C$ separately. Hereafter, $I_{C+}$ and $I_{C-}$ denote absolute values. The first crucial observation is that the symmetry of $I_{EVEN}$ and $I_{ODD}$ is similar in all three nanowires, determined solely by the angle $\theta$ between **H** and **I**, and is unrelated to the crystallographic axis. As shown in Figure 3, both $I_{EVEN}$ and $I_{ODD}$ can be well-fitted by functions of $\theta$ as follows:

$$I_{EVEN} = A\cos^4(\theta - 45) + B \qquad (1)$$

$$I_{ODD} = C\cos^4\theta + D\cos^4(\theta - \theta_0) \qquad (2)$$

where $A$, $B$, $C$, and $D$ represent fitting parameters (see Extended Data Table 1 for detailed fitting parameters). Remarkably, both $I_{EVEN}$ and $I_{ODD}$, which measure the absolute value of the superconducting critical current and its non-reciprocity, respectively, exhibit a surprisingly acute dependence on $\theta$ that can only be fitted by fourth-power cosine functions. This characteristic behaviour is typical of the magnetic chiral effect (MCE) observed in topological Dirac/Weyl semi-metals, originating from the chiral anomaly effect between 3D Dirac/Weyl cones with opposite chirality[21,22,27,47]. The $\cos^4\theta$-dependence has been theoretically predicted [47] and observed in the MCE of a TDS,

Na$_3$Bi[48], where the magnetoconductance varies as $\cos^4\theta$ at magnetic fields lower than 2 T. The very similar dependence on current and magnetic field directions between the critical current ($I_C$) in β-Sn/α-Sn nanowires and the magnetoconductance in TDSs strongly suggests that the TDS α-Sn areas play an important role in the superconductivity of the studied nanowires. It is noteworthy that although superconductivity has been realised in another TDS, Cd$_3$As$_2$, such a strong manifestation of TDS properties in the superconductivity has never been reported.

Additionally, other intriguing observations from Fig. 3 pertain to the symmetry of $I_{EVEN}$ and $I_{ODD}$. The critical current of β-Sn/α-Sn nanowires, measured by $I_{EVEN}$, exhibits significant two-fold anisotropy when **H** is rotated by 45° and 225° with respect to **I**. This reflects an anisotropic strength of the pairing potential of Cooper pairs, likely stemming from the distinct spin-orbit-momentum textures and DOS at $E_F$ when the configuration between **H** and **I** changes. Theoretical predictions have suggested anisotropic superconducting pairing potential and nodal superconducting gap structures for TDSs [20,49], which might be related to the observed anisotropic $I_{EVEN}$. Meanwhile, the non-reciprocal component $I_{ODD}$ also shows substantial enhancements when $\theta = 0°$ (**H** // **I**) and $\theta = 112°$, a unique and distinct feature from other cases of previous SDE. To explain the angular dependence of $I_{EVEN}$ and $I_{ODD}$, theoretical models accounting for the special spin-orbit-momentum texture in TDSs, particularly α-Sn, are strongly desired.

To gain insights into why the TDS properties of α-Sn strongly influence the superconductivity in the β-Sn nanowires, we explore the nanowire-width ($W$) dependence of the superconducting transport properties. As shown in Fig. 4a, in three nanowires with $W$ = 500, 1000, and 2000 nm, the values of $I_C$ in both positive and negative directions ($I_{C+}$, $I_{C-}$) increase linearly with $W$, as the same current density is required to reach the superconducting depairing limit. However, the non-reciprocal component $I_{ODD}$ exhibits weak dependence on $W$, suggesting that the SDE predominantly occurs at the edge of the β-Sn nanowires, precisely at the interfaces between β-Sn and α-Sn. According to the STEM results, the β-Sn nanowires consist of a disordered center part with a width $W_0$, formed by direct collisions with Ga ions, and disorder-free edge parts, formed by thermal diffusion on both edges (see Extended Data Figure 1). SEM images of the three nanowires in Fig. 4a confirm that $W \approx W_0 + 150$ nm, indicating that the thermally formed β-Sn region is approximately 75 nm wide on each side (see thermally formed β-Sn in Fig. 4b, orange coloured). Furthermore, the α-Sn regions facing the nanowire edges also become superconducting (SC) due to the proximity effect from the adjacent β-Sn (see proximitised SC α-Sn in Fig. 4b, green-yellow coloured). Based on the width dependence of $I_C$, we propose that the non-reciprocal superconducting transport occurs at these thermally formed β-Sn / proximitised SC α-Sn edge channels. This edge transport scenario is reasonable for the following three reasons: i) The better crystal quality and absence of ion-impacted damages in the edge regions guarantee stronger superconductivity at the edges compared to the center part. ii) Only at the β-Sn/α-Sn interface is the inversion symmetry broken, a prerequisite for non-reciprocal transport. iii) The observed superconducting transport in our nanowires closely resembles the MCE in TDS, which can only be explained by the contribution from adjacent superconducting α-Sn. Overall, our findings suggest that the non-reciprocal superconducting transport in the β-Sn nanowires is intimately related to the presence of high-quality α-Sn/β-Sn interfaces at the edges, highlighting the significance of these interfaces in enabling novel phenomena in topological superconductivity.

Another hallmark of the SDE, predicted in theoretical models[35,41,42,43,44], is the sign reversal of $\eta$ with increasing magnetic field strength $H$. Although this behaviour has been observed in a few recent experiments, it remained quite subtle[50,51]. As depicted in Fig. 4c, in our β-Sn/α-Sn nanowire NW0 ($W = 500$ nm), $I_{C+}$, $I_{C-}$, and $\eta$ exhibit strong and uniform oscillations with a period $\Delta H = 2400$ Oe, with $\eta$ changing sign at a half-period $\Delta H/2 = 1200$ Oe when a parallel **H** is applied. The theories on noncentrosymmetric superconductors typically explain the sign reversal of $\eta$ as a change in the nature of the helical superconducting state, but this is not accompanied by the strong oscillations in $I_C$[35,41,42,43]. Conversely, the oscillatory behaviour of $I_C$ with $H$ has been observed in Josephson junctions of a TDS consisting of Nb/Cd$_3$As$_2$/Nb when **H** is applied parallel to **I** across the junction[52]. This phenomenon was explained as an orbital effect when magnetic fluxes from **H** thread through a trajectory of the currents traveling between the bulk (Dirac cones) and surface (Fermi arcs) states of the TDS. Furthermore, in topological Dirac/Weyl semi-metals, the superconducting current can be carried separately in the bulk states and in the surface states[52,53], with the latter typically exhibiting stronger superconductivity due to higher mobility and robustness against disorder scattering. In our Sn nanowires, with a London penetration depth of β-Sn being 36 nm[54], the magnetic fluxes from a parallel **H** can penetrate the entire thickness (70 nm) of the β-Sn/α-Sn edge channels, assuming that the London penetration depth of the superconducting α-Sn (proximitised by the superconducting β-Sn) is the same as that of β-Sn. In this case, the value of $\Delta H$ = 2400 Oe corresponds to a magnetic flux quantum $\phi_0$ (=2.0×10$^{-15}$ Wb) threading through a cross-sectional area of a hypothetical Sn-based nanowire with a width $W^*$ = 150 nm and a film thickness of 70 nm in sample NW0. Although $W^*$ is much smaller than $W$ = 500 nm in NW0, it closely matches the width of the β-Sn/α-Sn edge channels, comprising the 75 nm-wide thermally formed β-Sn and the proximitised SC α-Sn with a typical width of ~100 nm. These dimensions thus suggest that the SDE occurs in the superconducting edge channels of α-Sn/β-Sn with a total width $W^*$ = 150 nm and a film thickness of 70 nm, as illustrated in Figure 4b. In this scenario, the discreet penetration of magnetic flux quanta leads to an oscillation with a period of $\phi_0$ ($\Delta H$ = 2400 Oe) and a sign reversal at $\phi_0/2$ ($\Delta H/2$ = 1200 Oe) in the SDE rectification ratio $\eta$, as predicted in a recent theory for SDE in a superconducting 1D chiral nanotube[44]. Furthermore, the oscillatory behaviour of $I_C$ and $\eta$ as a function of **H** suggests that the non-reciprocal superconducting transport primarily originates from the superconducting transport in the surface states of the TDS α-Sn, induced by a proximity effect from the β-Sn. Nevertheless, further theoretical investigations are undoubtedly needed to fully understand the mechanism of SDE demonstrated in our Sn-based nanowires.

In conclusion, using FIB, we have demonstrated a universal and cost-effective method for fabricating nanoscale Sn-based superconductor/TDS planar heterojunctions of arbitrary shapes. The combination of superconductivity and TDS properties led to observing a giant SDE in the β-Sn/α-Sn nanowires when **H** is parallel to **I**, a remarkable finding not reported before. The SDE's controllability by a magnetic field and its independence of the crystallographic axis, likely arising from a chiral anomaly in TDS α-Sn, offer non-volatile control when interfacing the structure with a ferromagnetic material and a high degree of freedom in embedding superconducting diodes in various designs for future quantum computing circuits. In a broader context, Sn-based superconductor/TDS heterostructures hold promise as a novel platform for topological superconductor devices and topological quantum circuitry, enabling nanoscale fabrication of various shapes through direct drawing with an ion beam.

**Methods**

*Sample growth*

The single-crystalline α-Sn thin film was epitaxially grown on an InSb (001) substrate using low-temperature molecular-beam epitaxy. First, a 100 nm-thick InSb buffer layer was grown on the InSb (001) substrate. Subsequently, the substrate holder was cooled to –10°C, and a 70 nm-thick α-Sn film was grown at a growth rate of 10 Å/min. The existence and quality of α-Sn were confirmed by observing the typical peak at 56.75° in $\omega - 2\theta$ measurements using an X-ray diffraction (XRD) measurement system, indicating the successful growth of single-crystalline α-Sn on the InSb substrate (see Extended Data Fig. 2).

*Device fabrication*

The β-Sn wire structures were fabricated by irradiating a Ga-focused ion beam using a V400ACE FIB system made by FEI Company Japan Ltd. An acceleration voltage of 30 kV and a current of 7.7 pA were used for the ion beam in all devices. The actual width $W$ of the β-Sn nanowires was measured using SEM, indicating that the β-Sn region induced by thermal diffusion is expected to be about 75 nm on each side. The β-Sn nanowire structures with widths of $W$ = 500, 1000, and 2000 nm were fabricated by irradiating FIB in regions with initial widths ($W_0$) of 350, 850, and 1850 nm, respectively. XPS measurements confirmed that no Ga remained in the irradiated Sn areas (data not shown).

*Transport measurements*

Direct current transport measurements were conducted using a Quantum Design Physical Property Measurement System. A sample holder equipped with a rotator was utilised, and the critical current was measured while incrementally rotating the sample in an in-plane magnetic field by 360° at 5° intervals.

*Critical current and non-reciprocal component of superconducting β-Sn wires*

The critical current is defined as the value just before the resistance increases when the current is swept from zero. Initially, we investigated the dependence of the $\eta$ value on the applied magnetic field $H$ when **H** is parallel to **I**. The critical currents of the β-Sn wires with different widths shown in Figure 4a and the in-plane magnetic field angle dependence of non-reciprocal superconductivity shown in Figure 3 were measured at a fixed magnetic field strength $H$ for each wire where $\eta$ reaches its maximum.

**REFERENCES**

1. Kitaev, A. Y. Unpaired Majorana fermions in quantum wires. *Physics-Uspekhi* **44**, 131–136 (2001).
2. Fu, L. & Kane, C. L. Superconducting proximity effect and Majorana fermions at the surface of a topological insulator. *Phys. Rev. Lett.* **100**, 096407 (2008).
3. Sarma, S. Das, Freedman, M. & Nayak, C. Majorana zero modes and topological quantum computation. *Npj Quantum Inf* **1**, 15001 (2015).
4. Sato, M. & Ando, Y. Topological superconductors: a review. *Reports on Progress in Physics* **80**, 076501 (2017).
5. Wang, M.-X. *et al.*, The coexistence of superconductivity and topological order in

the Bi$_2$Se$_3$ thin films. *Science* **336**, 52–55 (2012).

6. Wiedenmann, J. *et al.* 4π-periodic Josephson supercurrent in HgTe-based topological Josephson junctions. *Nature Commun.* **7**, 10303 (2016).

7. Huang, C. *et al.* Proximity-induced surface superconductivity in Dirac semimetal Cd$_3$As$_2$. *Nature Commun.* **10**, 2217 (2019).

8. Li, N. *et al.*, Gate modulation of anisotropic superconductivity in Al–Dirac semimetal Cd3As2 nanoplate–Al Josephson junctions. *Supercond Sci Technol* **35**, 044003 (2022).

9. Huang, C. *et al.* Inducing strong superconductivity in Wte$_2$ by a proximity effect. *ACS Nano* **12**, 7185–7196 (2018).

10. Li, Q. *et al.* Proximity-induced superconductivity with subgap anomaly in type II Weyl semi-metal Wte$_2$. *Nano Lett.* **18**, 7962–7968 (2018).

11. Das, A. *et al.* Zero-bias peaks and splitting in an Al-InAs nanowire topological superconductor as a signature of Majorana fermions. *Nature Phys.* **8**, 887–895 (2012).

12. Deng, M. T. *et al.*, Anomalous zero-bias conductance peak in a Nb-InSb nanowire-Nb hybrid device. *Nano Lett.* **12**, 6414–6419 (2012).

13. Leijnse, M. & Flensberg, K. Introduction to topological superconductivity and Majorana fermions. *Semicond Sci Technol* **27**, 124003 (2012).

14. Frolov, S. M., Manfra, M. J. & Sau, J. D. Topological superconductivity in hybrid devices. *Nature Phys.* **16**, 718–724 (2020).

15. Kezilebieke, S. *et al.* Topological superconductivity in a van der Waals heterostructure. *Nature* **588**, 424–428 (2020).

16. Anh, L. D. *et al.*, Elemental topological Dirac semimetal α-Sn with high quantum mobility. *Adv. Mater.* **33**, 2104645 (2021).

17. Wang, Z., Weng, H., Wu, Q., Dai, X. & Fang, Z. Three-dimensional Dirac semimetal and quantum transport in Cd$_3$As$_2$. *Phys. Rev. B* **88**, 125427 (2013).

18. Liu, Z. K. *et al.*, Discovery of a three-dimensional topological Dirac semimetal, Na$_3$Bi. *Science* **343**, 864-867 (2014).

19. Roy, B., Ghorashi, S. A. A., Foster, M. S. & Nevidomskyy, A. H., Topological superconductivity of spin-3/2 carriers in a three-dimensional doped Luttinger semimetal. *Phys. Rev. B* **99**, 054505 (2019).

20. Kobayashi, S. & Sato, M., Topological Superconductivity in Dirac Semimetals. *Phys. Rev. Lett.* **115**, 187001 (2015).

21. Wang, Z. *et al.* Dirac semimetal and topological phase transitions in $A_3Bi$ (A = Na, K, Rb). *Phys. Rev. B* **85**, 195320 (2012).

22. Feng, J. *et al.* Large linear magnetoresistance in Dirac semimetal $Cd_3As_2$ with Fermi surfaces close to the Dirac points. *Phys. Rev. B* **92**, 081306 (2015).

23. Ohtsubo, Y., Le Fèvre, P., Bertran, F., Taleb-Ibrahimi, A., Dirac cone with helical spin polarization in ultrathin α-Sn(001) films. *Phys. Rev. Lett.* **111**, 216401 (2013).

24. Barfuss, A., *et al.*, Elemental topological insulator with tunable Fermi level: Strained α-Sn on InSb(001). *Phys. Rev. Lett.* **111**, 157205 (2013).

25. Scholz, M. R. *et al.*, Topological surface state of α−Sn on InSb(001) as studied by photoemission. *Phys. Rev. B* **97**, 075101 (2018).

26. Rojas-Sánchez, J.-C. *et al.* Spin to charge conversion at room temperature by spin pumping into a new type of topological insulator: α -Sn films. *Phys. Rev. Lett.* **116**, 096602 (2016).

27. Uchida, M. *et al.* Quantum Hall states observed in thin films of Dirac semimetal Cd3As2. *Nature Commun.* **8**, 2274 (2017).

28. Burnell, G. *et al.* Planar superconductor-normal-superconductor Josephson junctions in $MgB_2$. *Appl. Phys. Lett.* **79**, 3464–3466 (2001).

29. Wu, C. H. *et al.* Fabrication and characterization of high- Tc $Yba_2Cu_3O_{7-x}$ nano SQUIDs made by focused ion beam milling. *Nanotechnology* **19**, 315304 (2008).

30. Wu, C.-H. *et al.* Fabrication and properties of high-Tc YBCO Josephson junction and squid with variable thickness bridges by focused ion beam. *IEEE Trans. Appl. Supercond.* **21**, 375–378 (2011).

31. Müller, B. *et al.* Josephson junctions and SQUIDs created by focused Helium-ion-beam irradiation of $Yba_2Cu_3O_7$. *Phys. Rev. Appl.* **11**, 044082 (2019).

32. Gál, N., Štrbík, V., Gaži, Š., Chromik, Š. & Talacko, M. Resistance anomalies at superconducting transition in multilayer N/S/F/S/N heterostructures. *J. Supercond. Nov. Magn.* **32**, 213–217 (2019).

33. Ando, F. *et al.* Observation of superconducting diode effect. *Nature* **584**, 373–376 (2020).

34. Jiang, K. & Hu, J. Superconducting diode effects. *Nature Phys.* **18**, 1145–1146 (2022).

35. Yuan, N. F. Q. & Fu, L. Supercurrent diode effect and finite-momentum superconductors. *Proc. Natl. Acad. Sci.* **119**, e2119548119 (2022).

36. Lin, J.-X. *et al.* Zero-field superconducting diode effect in small-twist-angle trilayer graphene. *Nature Phys* **18**, 1221–1227 (2022).

37. Pal, B. *et al.* Josephson diode effect from Cooper pair momentum in a topological semimetal. *Nature Phys* **18**, 1228–1233 (2022).

38. Baumgartner, C. *et al.* Supercurrent rectification and magnetochiral effects in symmetric Josephson junctions. *Nature Nanotechnol.* **17**, 39–44 (2022).

39. Bauriedl, L. *et al.* Supercurrent diode effect and magnetochiral anisotropy in few-layer NbSe$_2$. *Nature Commun.* **13**, 4266 (2022).

40. Wu, H. *et al.* The field-free Josephson diode in a van der Waals heterostructure. *Nature* **604**, 653–656 (2022).

41. Daido, A., Ikeda, Y. & Yanase, Y. Intrinsic superconducting diode effect. *Phys. Rev. Lett.* **128**, 037001 (2022).

42. Ilic, S. & Bergeret, F. S., Theory of the supercurrent diode effect in Rashba superconductors with arbitrary disorder. *Phys. Rev. Lett.* **128**, 177001 (2022).

43. He, J. J., Tanaka, Y. & Nagaosa, N. A phenomenological theory of superconductor diodes. *New J. Phys.* **24**, 053014 (2022).

44. He, J. J., Tanaka, Y. & Nagaosa, N. The supercurrent diode effect and nonreciprocal paraconductivity due to the chiral structure of nanotubes. *Nature Commun.* **14**, 3330 (2023).

45. Hou, Y. *et al.* Ubiquitous Superconducting Diode Effect in Superconductor Thin Films, arXiv:2205.09276v5 (2022).

46. Suri, D. *et al.* Non-reciprocity of vortex-limited critical current in conventional superconducting micro-bridges. *Appl. Phys. Lett.* **121**, 102601 (2022).

47. Burkov, A.A. & Kim, Y. B., Z2 and chiral anomalies in topological Dirac semimetals. Phys. Rev. Lett. **117**, 136602 (2016).

48. Jun, X. *et al.* Evidence for the chiral anomaly in the Dirac semimetal Na3Bi. *Science* **350**, 413–416 (2015).

49. Hashimoto, T., Kobayashi, S., Tanaka, Y., & Sato, M., Superconductivity in doped Dirac semimetals. *Phys. Rev. B* **94**, 014510 (2016).

50. Kawarazaki, R. *et al.*, Magnetic-field-induced polarity oscillation of superconducting diode effect, *Appl. Phys. Express* **15**, 113001 (2022).

51. Margineda, D., Crippa, A., Strambini, E., Fukaya, Y., Mercaldo, M. T., Cuoco, M., & Giazotto, F., Sign reversal diode effect in superconducting Dayem nanobridges,

arXiv:2306.00193 (2023).

52. Li, C. Z. *et al.* Fermi-arc supercurrent oscillations in Dirac semimetal Josephson junctions. *Nature Commun* **11**, 1150 (2020).

53. Kuibarov, A. *et al.* Superconducting Arcs, *arXiv*:2035.02900 (2023).

54. Kozhevnikov, V. *et al.*, Nonlocal effect and dimensions of Cooper pairs measured by low-energy muons and polarized neutrons in type-I superconductors. *Phys. Rev. B* **87**, 104508 (2013).

arXiv:2306.00193 (2023).

52. Li, C. Z. *et al.* Fermi-arc supercurrent oscillations in Dirac semimetal Josephson junctions. *Nature Commun* **11**, 1150 (2020).

53. Kuibarov, A. *et al.* Superconducting Arcs, *arXiv*:2035.02900 (2023).

54. Kozhevnikov, V. *et al.*, Nonlocal effect and dimensions of Cooper pairs measured by low-energy muons and polarized neutrons in type-I superconductors. *Phys. Rev. B* **87**, 104508 (2013).

**Data and materials availability:** The data that support the findings of this study are available from the corresponding authors upon reasonable request.

**Acknowledgments:**
This work was partly supported by Grants-in-Aid for Scientific Research (Grants No. 20H05650 and No. 23K17324), CREST program (JPMJCR1777) and PRESTO Program (JPMJPR19LB) of JST, UTEC-UTokyo FSI, Murata Science Foundation and Spintronics Research Network of Japan (Spin-RNJ). A part of this work was supported by "Advanced Research Infrastructure for Materials and Nanotechnology in Japan (ARIM)" of the Ministry of Education, Culture, Sports, Science and Technology (MEXT).

**Author contributions:**
L.D.A and M.T. supervised the study. K.I. and L.D.A. conceived and designed the experiments. T.H. performed the growth and characterisation of the α-Sn thin film. K.Inagaki and M.K. carried out XPS measurements. K.I. and L.D.A. fabricated β-Sn wire devices using FIB and measured the transport properties. K.I. and L.D.A carried out the data analysis. K. I, L. D. A. and M. T. wrote the manuscript. All authors thoroughly discussed and commented on the manuscript.

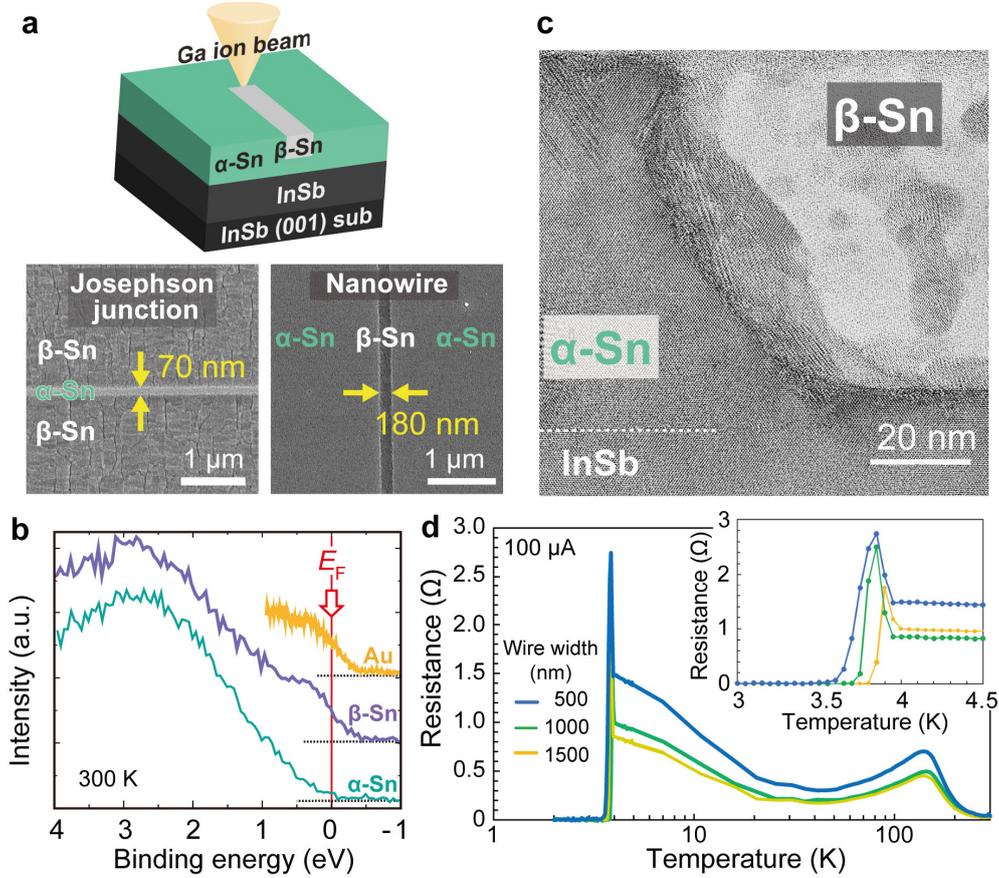

**Figure 1. Phase transition from topological Dirac semimetal α-Sn to superconducting metal β-Sn induced by irradiation of a Ga-focused ion beam (FIB).** **a,** Schematic image (top) and top-view SEM images of α-Sn/β-Sn planar nanostructures fabricated by FIB irradiation. A β-Sn/α-Sn/β-Sn Josephson junction structure with a α-Sn width of 70 nm (bottom left) and a 180 nm-wide β-Sn nanowire embedded in α-Sn (bottom right) are successfully fabricated (see Methods). **b,** XPS measurements of the valence band spectrum of α-Sn and β-Sn regions, compared with an Au reference film. A large DOS is detected at the Fermi edge of the irradiated Sn, reflecting the metallic behaviour of β-Sn. **c,** Cross-sectional STEM lattice image of the interface region between α-Sn/β-Sn prepared by FIB irradiation. The as-grown α-Sn retains its diamond-type crystal structure, while the FIB-irradiated region changes to polycrystalline β-Sn. An abrupt interface between α-Sn/β-Sn region is obtained. **d,** Temperature dependence of the resistance of the β-Sn wires with various wire widths. All the wires undergo a superconducting transition below 4 K.

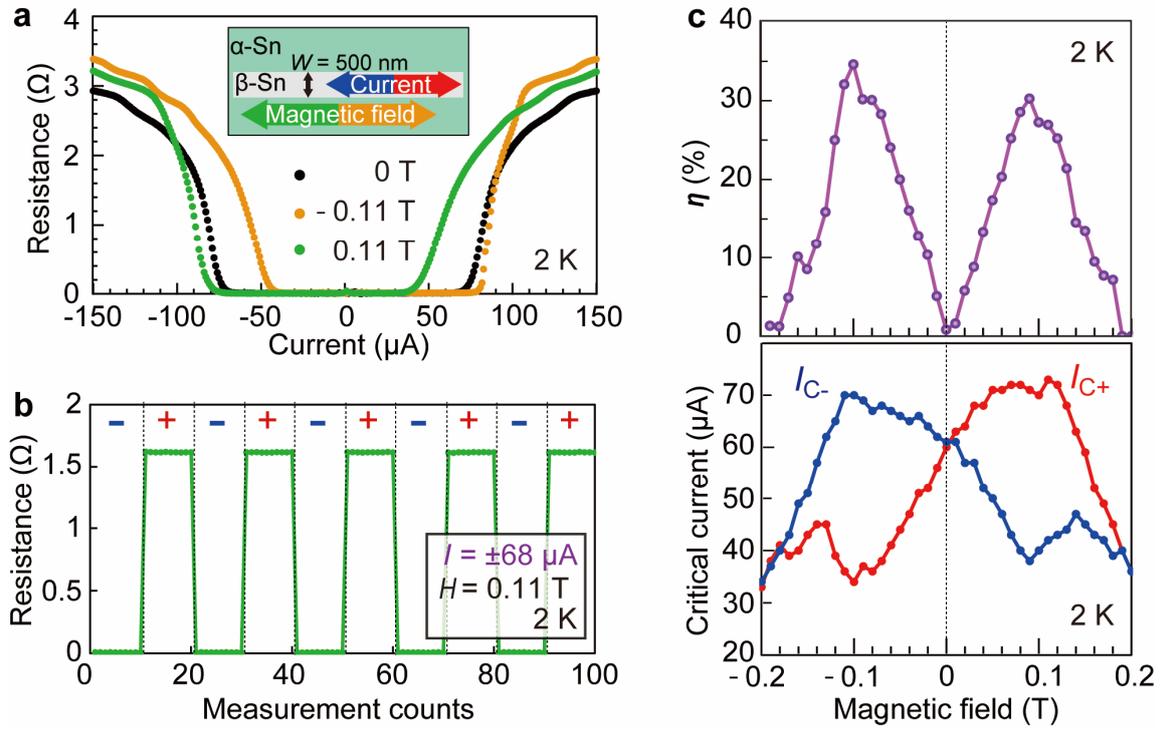

**Figure 2. Observation of giant non-reciprocal superconductivity. a,** Current dependence of the resistance of a β-Sn/α-Sn nanowire with a width $W$ = 500 nm when a magnetic field **H** is applied parallel (orange) and antiparallel (green) to the current **I**. **b,** Switching between the superconducting and normal-conducting states by changing the direction of the applied current between ±68 μA, while a magnetic field $H$ = 0.11 T is applied parallel to current direction. **c,** Diode rectification ratio $\eta$ and critical current $I_C$ as a function of the magnetic field strength with **H** // **I**. $\eta$ reaches a maximum value of 35% when $H$ = 0.1 T. All were measured at 2 K.

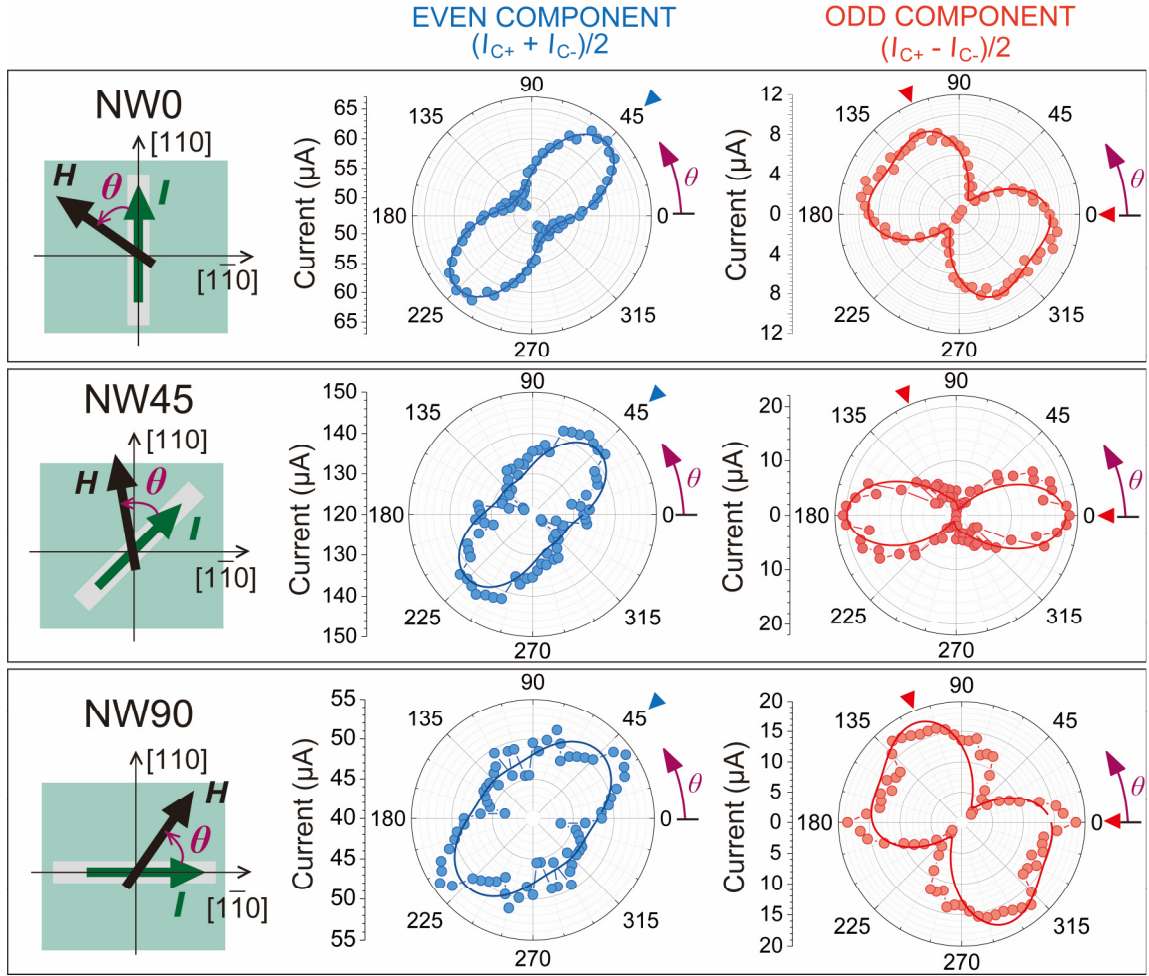

**Figure 3. In-plane magnetic-field angular dependence of the non-reciprocal critical currents in three β-Sn/α-Sn nanowires ($W$ = 500 nm, namely NW0, NW45, NW90) oriented along three different crystallographic axes of [110], [100] and $[1\bar{1}0]$, respectively.** The magnetic-field angular dependence of the even component $I_{EVEN}$ (blue data points) and the odd component $I_{ODD}$ (red data points) of the critical currents show different symmetries, which can be fitted as functions of $\theta$, the angle between **H** and **I**, using equations (1) and (2). Blue and red triangles point to the two-fold symmetry axes of the even and odd components, respectively. Fixed values of $H$ (= 0.056, 0.185, 0.110 T for NW0, NW45, and NW90, respectively) are applied. All are measured at 2 K.

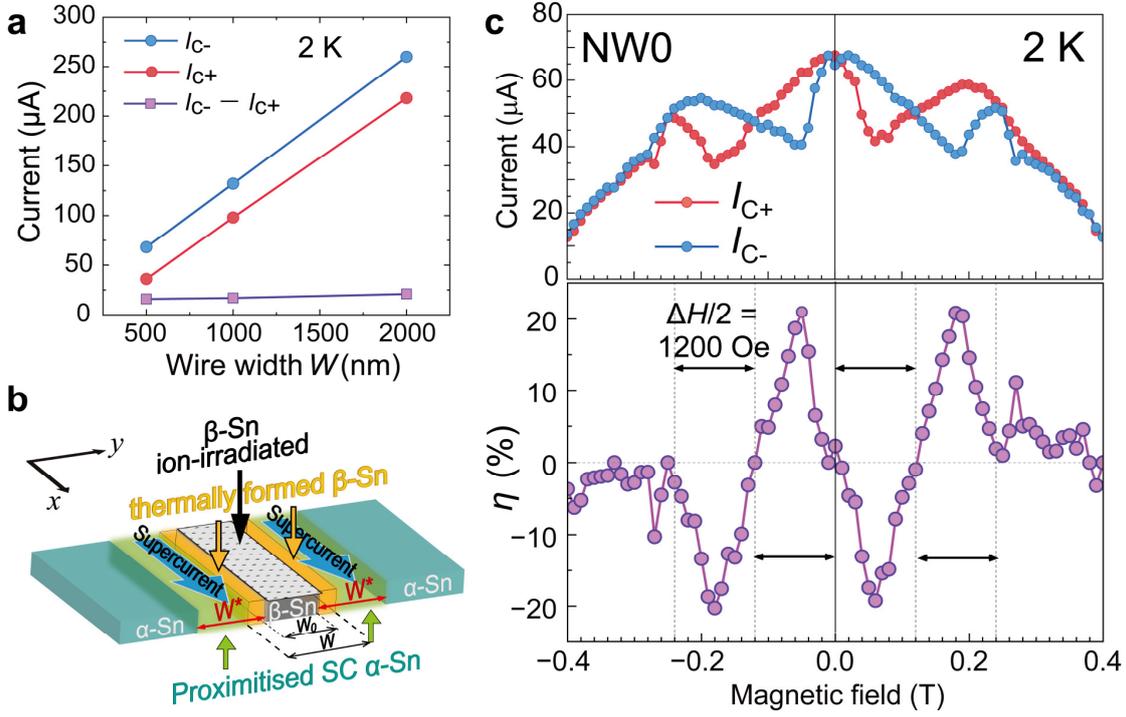

**Figure 4. Width and magnetic field dependence of non-reciprocal superconductivity in β-Sn/α-Sn nanowires. a,** Width dependence of the critical currents in the positive direction $I_{C+}$ and negative direction $I_{C-}$, and their difference, $I_{C+} - I_{C-}$. **b,** Schematic view of the β-Sn/α-Sn planar structure deduced from the results in **a**. The β-Sn nanowires contain a disordered center part (grey-coloured) with a width $W_0$ formed by direct collisions with the Ga ions and disorder-free edge regions (orange-coloured) formed by thermal diffusion in both edges. Furthermore, the α-Sn areas that face the nanowire edges also become superconducting due to a proximity effect from β-Sn (green-yellow coloured). The total thickness of the thermally formed β-Sn and the proximitised superconducting (SC) α-Sn on each side is $W^*$, which is about 150 nm. **c,** Dependence of the critical currents in the positive direction $I_{C+}$ and in the negative direction $I_{C-}$, and $\eta$ on the magnetic field strength in NW0 when **H** is applied parallel to **I**. Strong and uniform oscillations with a period $\Delta H$ = 2400 Oe are observed, and $\eta$ changes its sign at the half-period $\Delta H/2$ = 1200 Oe. All are measured at 2 K.

**Extended Data Table 1.** Parameters obtained from least-square fitting using equations (1) and (2) to the in-plane magnetic-field angular dependence of non-reciprocal critical currents in NW0, NW45, and NW90, shown in Figure 3. The error bars of the parameters are given by the standard errors of the fitting process.

|  | NW0 | NW45 | NW90 |
|---|---|---|---|
| $A$ (μA) | 13.9±0.4 | 14.8±0.8 | 4.9±0.5 |
| $B$ (μA) | 49.1±0.2 | 126.5±0.5 | 46.8±0.3 |
| $C$ (μA) | 8.8±0.2 | 19.7±0.6 | 14.1±0.7 |
| $D$ (μA) | 8.8±0.2 | 3.2±0.6 | 17.6±0.7 |
| $\theta_0$ (deg) | 112.4±0.7 | 112 (fixed) | 112 (fixed) |

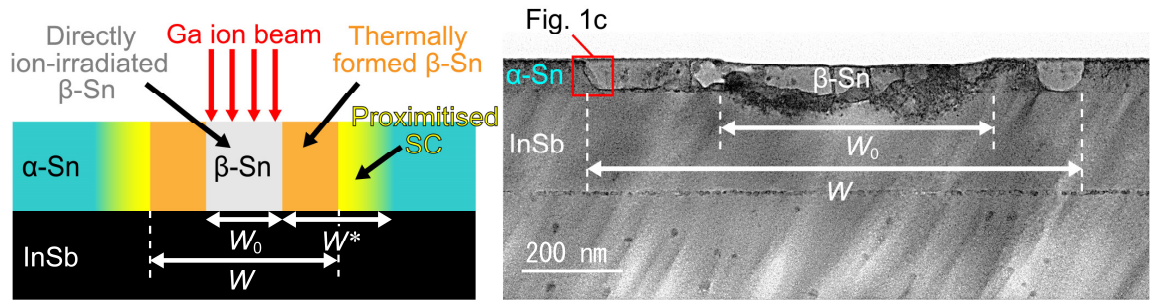

**Supplementary Fig.1. Schematic structure and cross-sectional STEM image of an FIB-induced β-Sn wire embedded in an α-Sn thin film.** As shown in the schematic illustration (left), the actual β-Sn wire with a width $W$ consists of a center region with a width $W_0$ (grey coloured), where α-Sn is directly irradiated by the Ga ion beam, and two disorder-free β-Sn regions (orange coloured) on both sides that are formed due to thermal diffusion. Crystal defects and boundaries are observed in the center FIB-irradiated region, which are polycrystalline β-Sn with multiple grains. Furthermore, there are superconducting (SC) α-Sn regions (green-yellow coloured) due to a proximity effect from β-Sn. The total thickness of the thermally formed β-Sn and the proximitised SC α-Sn on each side is named $W^*$. In this specific sample, the width of the thermally formed β-Sn regions are wider than 75 nm and differs between the left and right sides, as shown in the STEM image on the right. A possible reason for this might be that the thermally formed β-Sn regions are gradually widened after so many transport measurement cycles. When the current is above the superconducting critical current, Joule heating is generated in the normal-conducting state, which depends on the density of defects and is yielded asymmetrically on the left and right sides. This Joule heating may cause the widening of the thermally formed β-Sn parts, as observed in this STEM sample.

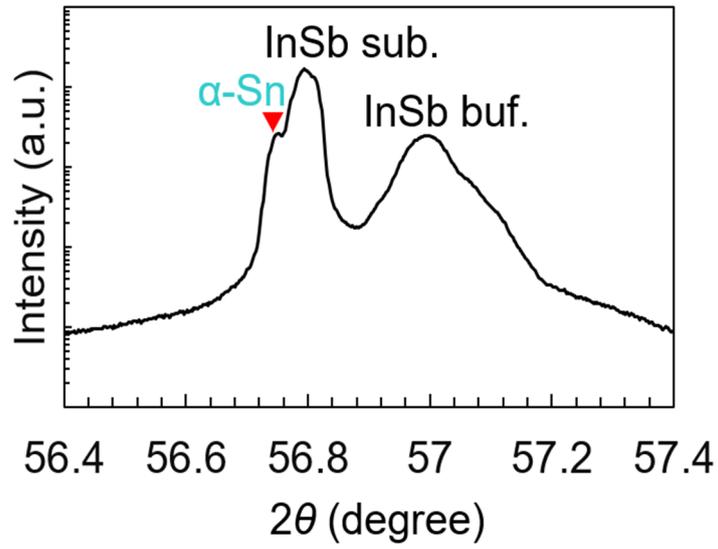

**Supplementary Fig.2. X-ray diffraction (XRD) *ω-2θ* measurement of the α-Sn thin film grown on an InSb (001) substrate using XRD.** The typical peak at $2\theta =$ 56.75° indicates that an epitaxial single-crystalline α-Sn thin film is successfully grown on the InSb substrate.